\documentclass[preprint,12pt]{aastex}
\usepackage{apjfonts}
\usepackage[dvips]{color}
\usepackage{amssymb, amsmath, amsbsy, epsfig, epsf, slashbox}
\newcommand{\lum}{erg~s\ensuremath{^{-1}}}
\newcommand{\lbol}{\ensuremath{L\mathrm{_{bol}}}}

\newcommand{\ledd}{\ensuremath{L\mathrm{_{Edd}}}}
\newcommand{\lratio}{\ensuremath{L/\ledd}}

\newcommand{\lfive}{\ensuremath{\lambda L_{\lambda}(5100}~\AA)}

\newcommand{\msun}{\ensuremath{M_{\odot}}}
\newcommand{\kms}{\ensuremath{\mathrm{km~s^{-1}}}}
\newcommand{\mbh}{\ensuremath{M_\mathrm{BH}}}

\newcommand{\rs}{\ensuremath{r_{\rm \scriptscriptstyle S}}}
\newcommand{\pnull}{\ensuremath{P_{\mathrm{null}}}}

\newcommand{\ha}{{\rm H\ensuremath{\alpha}}}
\newcommand{\hb}{{\rm H\ensuremath{\beta}}}

\newcommand{\sii}{[S\,{\footnotesize II}]}
\newcommand{\oiii}{{\rm [O\,{\footnotesize III}]}}

\newcommand{\feii}{Fe\,{\footnotesize II}}

\newcommand{\mgii}{Mg\,{\footnotesize II}}

\newcommand{\civ}{C\,{\footnotesize IV}}

\def\lax{{$\mathrel{\hbox{\rlap{\hbox{\lower4pt\hbox{$\sim$}}}\hbox{$<$}}}$}}
\def\gax{{$\mathrel{\hbox{\rlap{\hbox{\lower4pt\hbox{$\sim$}}}\hbox{$>$}}}$}}

\def\kms{\hbox{km$\,$s$^{-1}$}}

\begin{document}
%\begin{CJK}{GBK}{song}

\slugcomment{To appear in {\it The Astrophysical Journal}}
%\slugcomment{\bf \it \small \color{red} Notice: be sure to toggle
%off any auto-Wrap function of your Tex editor, thus to enable us to
%compare different versions by UltraEdit.}

\title{The Blueshifting and Baldwin effects for the \oiii\,$\lambda5007$
Emission Line in Type~1 Active Galactic Nuclei}

\author{
Kai~Zhang\altaffilmark{1}, Xiao-Bo~Dong\altaffilmark{1,2},
Ting-Gui~Wang\altaffilmark{1}, and C.~Martin~Gaskell\altaffilmark{3}}

\altaffiltext{1}{Key laboratory for Research in Galaxies and Cosmology,
Department of Astronomy, The University of Sciences and Technology of China,
Chinese Academy of Sciences, Hefei, Anhui 230026, China; \\
zkdtc@mail.ustc.edu.cn;  xbdong@ustc.edu.cn;  twang@ustc.edu.cn}
\altaffiltext{2}{The Observatories of the Carnegie Institution for
Science, 813 Santa Barbara Street, Pasadena, CA 91101, USA}
\altaffiltext{3}{Departamento de F\'isica y Astronom\'ia, Facultad de Ciencias,
Universidad de Valpara\'iso, Av. Gran Breta\~na 1111, Valpara\'iso, Chile.
martin.gaskell@uv.cl}

%\email{zkdtc@mail.ustc.edu.cn}
\shorttitle{Blueshifting and Baldwin effects for \oiii\,$\lambda5007$}
\shortauthors{Zhang et al.}

\begin{abstract}

We use homogeneous samples of radio-quiet Seyfert 1 galaxies and QSOs
selected from the Sloan Digital Sky Survey
to investigate the connection between the velocity shift and the
equivalent width (EW) of the \oiii\,$\lambda5007$ emission line,
and their correlations with physical parameters of active galactic nuclei (AGNs).
We find a significant and negative correlation between the EW
of the core component, EW(core), and the blueshift of either the
core (the peak), the wing, or the total profile of \oiii\ emission;
it is fairly strong for the blueshift of the total profile particularly.
However, both quantities (EW and velocity shift) generally
have only weak, if any,
correlations with fundamental AGN parameters such as the nuclear continuum
luminosity at 5100~\AA\ ($L_{5100}$), black hole mass (\mbh),
and the Eddington ratio (\lratio); these correlations include
the classical Baldwin effect of EW(core),
an inverse Baldwin effect of EW(wing), and the relationship between velocity
shifts and \lratio. Our findings suggest that both the large
object-to-object variation in the strength of \oiii\ emission and the
blueshift--EW(core) connection are not governed primarily by fundamental
AGN parameters such as $L_{5100}$, \mbh\ and \lratio.
We propose that the ISM conditions of the host galaxies
play a major role instead in
the diversity of the \oiii\ properties in active galaxies.  This suggests
that the use of \oiii\,$\lambda5007$ luminosity as proxy of
AGN luminosity does not depend strongly on the above-mentioned
fundamental AGN parameters.
\end{abstract}

\keywords{galaxies: active - galaxies: Seyfert - quasars: emission lines }

\section{Introduction}

In terms of size, the narrow-line region (NLR) of active galactic
nuclei (AGNs; including Seyfert galaxies and quasars) represents the link
between the inner structure of the AGN---the dusty torus and
the broad-line region (BLR)---and the interstellar medium in
the host galaxy.  The NLR bridges the region dominated by the
gravitational field of the central massive black hole and
the region dominated by the gravitational potential of the bulge stars
of the host galaxy.
To gain a complete understanding of the AGN phenomenon, the
role of the NLR therefore needs to be understood.
Unlike the BLR, the NLR, or at least the outer parts of the
NLR, can be spatially resolved in nearby AGNs.  Despite this,
however, the NLR remains remarkably poorly understood and there
are many unanswered questions about the structure and kinematics
of the NLR and its relationship to the BLR/torus.

\oiii\,$\lambda5007$ is one of the most important NLR lines.
Its equivalent width (EW) in the AGN ensemble varies
dramatically by a factor of $> 300$ (from $<0.5$ to 157 \AA; Baskin \& Laor 2005b),
one of the things showing the most object-to-object variation in AGNs.
It is a dominant variable in the set of correlations making up the
first principal component (commonly called ``Eigenvector~1''; EV1)
in the principal component analysis of QSO properties of Boroson \& Green (1992);
EV1 is believed to be linked to certain fundamental parameters of
the accretion process. \oiii\ luminosity is commonly used
as a surrogate for the bolometric luminosity (e.g., Heckman et al. 2004) and its
Doppler width as a surrogate for the stellar velocity dispersion
($\sigma_\star$) of the host galaxy
(e.g., Nelson \& Whittle 1996, Wang \& Lu 2001, Komossa \& Xu 2007, Gaskell 2009a).

Although the spatial distribution of \oiii-emitting gas in the NLR of AGNs is
rather inhomogeneous (see, e.g., Das et al. 2006, Crenshaw et al. 2010), there
are some global regularities in its strength and kinematics.
Its EW was found to correlate negatively, albeit with a large scatter,
with the continuum luminosity (Grindlay et al. 1980, Steiner 1981,
Dietrich et al. 2002, Kovacevic et al. 2010;  cf. Table~3 of Dong et al. 2011).
This is similar to the famous ``Baldwin effect'' (Baldwin 1977, Baldwin et al. 1978)
first discovered for the broad \civ\,$\lambda1549$ line.
The \oiii\ line profile can be regarded as consisting of
two components: a narrow component (hereinafter, the `core')
with a velocity close to the systemic redshift of the host
galaxy and a low, broad base that is usually
blueshifted (hereinafter, the `wing'). There is mounting evidence that the
core component is a good tracer of the gravitational potential of
the host galaxy bulge (see, e.g., Gaskell 2009a and references therein)
while the wing component arises from AGN outflows
(see, e.g., Crenshaw et al. 2010, Komossa \& Xu 2007 and references therein).
Scattering could also be a contributor to broadening of the wings (Gaskell \& Goosmann 2008).

Besides these general global trends in the
AGN ensemble, there are some interesting peculiar \oiii\ phenomena
that have been discovered in
some particular AGN subclasses. For instance, in narrow-line Seyfert
1s (NLS1s) that have broad \hb\ line with full width at half maximum
(FWHM) $< 2000$ \kms\ (the ``population A'' of the Eigenvector~1
domain, Zamanov et al. [2002]; see Xu \& Komossa [2010] for a recent review),
\oiii\,$\lambda5007$ is generally very
weak.  Moveover, in a small fraction of NLS1s, the whole
\oiii\,$\lambda5007$ profile has a bulk velocity shift of $\gtrsim 250$
\kms\ blueward (the so-called ``blue outliers'', Zamanov et al. 2002, Boroson
2005, Zhou et al. 2006, Komossa et al. 2008) and
there seems to be no non-blueshifted
component from a canonical NLR as seen in low-ionization lines
(Komossa et al. 2008).
Komossa \& Xu (2007) found that
although the width of the core component of \oiii\,$\lambda5007$ is generally
a good surrogate for $\sigma_\star$, this however does not hold in ``blue outliers''.

It is an interesting question whether or not these various regularities
and phenomena can be unified and this has been looked at by previous
authors. For instance, Zamanov et al. (2002) suggested that ``blue
outliers'' with extreme blue wing components originate in strong AGN outflows,
and moreover, are associated with the same outflows as the blue-shifted component
of the broad \civ\,$\lambda 1549$ emission line. The origin of the blueshifting of
high-ionization broad emission lines such as \civ\,$\lambda 1549$ remains
a subject of debate. There is general agreement that there has to be
a radial component of motion and some opacity source. Gaskell (1982) proposed
that the blueshifting was due to an outflow of the high-ionization gas and
obscuration by the accretion disk.  While this has remained the
most popular model, it has the major problem that outflows of the
dominant line-emitting gas seem to be strongly ruled out by
velocity-resolved reverberation mapping
(Gaskell 1988, Koratkar \& Gaskell 1989, see Gaskell 2009b for a review).
Gaskell \& Goosmann (2008) therefore argue that the blueshifting is
due instead to scattering of photons by material with a net inward
motion as suggested by Corbin (1990).

Observationally, the \civ\,$\lambda 1549$ Baldwin effect and blueshifting are
related in the sense that
higher-luminosity AGNs have both lower \civ\ EWs and larger blueshifts
(Corbin 1991, 1992).  This is further supported
by the study of Richards et al. (2002).  If this is also true for
\oiii\,$\lambda5007$, we will gain important insights
into the origin and kinematics of the NLR.  It might point to
similar underlying causes for the two effects in both the BLR and NLR.
Meanwhile, understanding these effects should enable us to calibrate better the
Doppler widths and luminosities of \oiii\,$\lambda5007$ to serve as
proxies for the bulge gravitational potential and AGN bolometric luminosity.

Motivated by these considerations, we present here the results of a
systematic study of the velocity shift of \oiii\,$\lambda5007$ emission
line, its connection to the Baldwin effect, and connection to
possible physical drivers.
This study takes advantage of the unprecedented spectroscopic data from the
Sloan Digital Sky Survey (SDSS; York et al. 2000).
Throughout this paper, we use a cosmology with
$H_{\rm 0}$ = 70 km\,s$^{-1}$\,Mpc$^{-1}$, $\Omega_{\rm m}$ = 0.3, and
$\Omega_{\rm \Lambda}$ = 0.7.

\section{Sample and Measurements}

\subsection{Samples}

Type~1 AGNs are chosen from the sample of Seyfert 1 galaxies and quasars
of Dong et al. (2011) taken from
the Fourth Data Release (Adelman-McCarthy et al. 2006) of the SDSS.
We select type~1 AGNs having the
\oiii\,$\lambda5007$ emission line in the SDSS bandpass and with
continuum and emission lines suffering only minimally from contamination
by host galaxy starlight. To ensure reliable analysis of the \oiii\,$\lambda5007$
profile, we also require high spectral
quality in the \oiii\,$\lambda5007$ region.
Our primary goal is to study the connection between velocity shifts
and EWs. To ensure the reliable analysis of the velocity shift of \oiii,
we first construct a basic sample by taking the
\sii\,$\lambda\lambda 6716,6731$ doublet lines as
the fiducial reference for the systematic redshift.
\sii\ is free from broad line contamination
so using it is more accurate than using narrow \hb\ (see Section 2.2).

The final criteria used for the basic sample are therefore:
(a) $z \leq 0.3$, to ensure that \sii\ emission lines, as well as
\oiii\,$\lambda5007$ and broad \hb, are in the bandpass;
(b) a median signal-to-noise ratio (S/N) $\geq 10$ per pixel for the
optical spectra, and particularly $\geq 15$ in the \oiii\ region
(4995--5020 \AA);
(c) the S/N of either \sii\ $\lambda 6716$ or $\lambda 6731$ line
$\geq 3$ to ensure \sii\ to be a reliable reference for velocity shift;
and (d) weak stellar absorption features, such that the
rest-frame EWs of Ca\,K (3934 \AA), Ca\,H~+~H$\epsilon$ (3970 \AA),
and H$\delta$ (4102 \AA) absorption features are undetected at
$< 2\,\sigma$ significance. The last criterion ensures the measurement
of the AGN luminosity and the emission-line EWs suffering little
from the contamination of the host galaxy starlight
(see the Appendix of Dong et al. 2011).

After removing duplications and sources with too many bad pixels in
the \hb\ + \oiii\ region, we obtain 565 type~1 AGNs.
Radio jets may enhance \oiii\ emission (Labiano 2008) and thus
possibly affect the results. We therefore use the method of Lu
et al. (2007) to exclude sources detected by the FIRST
(Faint Images of the Radio Sky at Twenty Centimeters) survey
by matching with FIRST catalog (Becker et al. 1995). After this, our basic
sample consist of 383 sources.

Because of the redshift cutoff of $z \leq 0.3$ (to ensure the presence of
\sii\ as the redshift reference), the dynamic ranges of AGN parameters
is somewhat restricted (see Subsection 3.2). So in the investigation
concerning \oiii\ EWs only,
we further define an extended sample by relaxing both the redshift cutoff and
the S/N requirement for \sii.  The extended sample comprises
1951 radio-quiet type~1 AGNs at $z<0.8$ culled from the Dong et al. (2011) sample,
with a median S/N $\gtrsim 15$ pixel$^{-1}$ in the \oiii\ region.

\subsection{Analysis of spectra}

Data analysis methods are as described in detail in Dong et al. (2011).
We only provide a brief description here. Following Dong et al. (2008),
we fit simultaneously the AGN featureless continuum, the
\feii\ multiplets, and other emission lines using a code based on
the MPFIT package (Markwardt 2009), and if necessary we refine the
emission-line fitting with the continuum subtracted spectra
using a code described in detailed in Dong et al. (2005).
The AGN continua are modeled by local power-laws
for 4200--5600 \AA\ and the \ha\ regions. The \feii\ emission is modeled
with two separate sets of templates in analytical forms, one
for the broad-line system and the other for the narrow-line system,
using the identification and measurement of \feii\ lines in I~Zw~1
from V\'eron-Cetty et al. (2004), as listed in their Tables~A1 and
A2.  Within each system, we assume that the respective sets of \feii\ lines
have no relative velocity shifts and have the same relative
strengths as those in I~Zw~1. Broad \feii\ lines are assumed to have
the same profile as broad \hb, while each narrow \feii\ line is
modeled with a Gaussian. During the fitting, the normalization and
redshift of every system are set to be free parameters. Broad
Hydrogen Balmer lines are fitted with as many Gaussians as is
statistically justified.

All narrow emission lines are fitted with a
single Gaussian, except for the \oiii\,$\lambda \lambda4959, 5007$
doublet lines. Each line of the doublet is modeled with two
Gaussians, one accounting for the line core and the other for a
possible blue wing, as seen in many objects. The Gaussian that has a
smaller velocity shift relative to \sii, whether
blueward or redward, is taken to be the core component. It also turns out
to be narrower than the other Gaussian component (if the other one was
present).  The velocity shifts of the core component with respect to \sii\
in the basic sample have
a mean of $-$47 \kms\ (negative values denoting blueshifts)
and a standard deviation of 72 \kms,
while the mean shift of the wing component is $-$225 \kms\ with a standard deviation
of 240 \kms, respectively.
On average the core components comprise 54 per cent of the total
emission, with a standard deviation of 0.16~dex.
We list in Table~1 the fluxes, EWs and velocity shifts of
the core and wing components and the whole profile of \oiii\,$\lambda5007$
for all the objects in the basic sample.
The velocity shifts are calculated by assuming that the \sii\ doublet lines
are located at the systematic redshift.
In Table~2 we list the fluxes and EWs of the core and wing components of
\oiii\,$\lambda5007$ for all the objects at higher redshifts that are thus
not in the basic sample but in the extended sample.
The other spectral parameters of this
sample are given in Dong et al. (2011).%
\footnote{Also available online at
http://staff.ustc.edu.cn/\~{ }xbdong/Data\_Release/ell\_effect/ \,.}
An example is illustrated in Fig.~1 of fitting the
\oiii\,$\lambda5007$ line profile of SDSS spectrum of J080131.58+354436.4
(i.e., the decomposition of the core and wing components).

In the basic sample the velocity shifts of narrow \hb\
with respect to \sii\ have a mean of $-21$ \kms\
and a standard deviation of 82 \kms.
Such a scatter is a bit larger than that of the core component of \oiii.
So if we use narrow \hb\ as the fiducial reference for
the systematic redshift to study the velocity shift of \oiii,
it would smear the correlations of interest in this study.
This should be especially true for the sources at relatively higher redshifts
in the extended sample. Those sources have lower spectral S/N
and a weaker narrow \hb\ component than the sources in the basic sample,
thus the deblending of narrow \hb\ from the broad component has
a larger uncertainty. The use of narrow \hb\ is checked in Section 3.

We notice that there is usually a prominent \feii\ emission feature
immediately redward of \oiii\,$\lambda5007$, which is dominated by
\feii\,42 $\lambda5018$ and [\feii]\,20F $\lambda5020$. To ensure
that the \oiii\ measurements suffer little from poor subtraction of
this \feii\ feature, we visually inspect the fits for the entire
extended sample and fine-tune the fits for some objects by carefully matching
the model to this local feature.  In such objects, the relative strengths
of \feii\ lines are significantly different from those of I~Zw~1 on which
the \feii\ template data are based.

\section{Results}

\subsection{The connection between the blueshifting and Baldwin effects
for \oiii\,$\lambda5007$}

We explore the connection between the velocity shift and
strength of \oiii\,$\lambda 5007$ in the basic sample.
We calculate the velocity shifts of the core and wing components
and the centroid (i.e., intensity-weighted average) of the whole profile
(\,$\Delta v$(core), $\Delta v$(wing), and $\Delta v$(centroid),
respectively) with respect to \sii.
The line strength is characterized by the rest-frame EW,
as measured by dividing the emission line flux by the underlying AGN continuum
flux density at the central wavelength,
for the core and wing components and the whole \oiii\ emission, respectively.
We perform bivariate Spearman rank correlations to test the relationship
between these kinds of velocity shifts and EWs.
This method tests for not only a linear relation but a monotonic one.
The results are summarized in Table~3,
where we report the Spearman coefficient (\rs) and the probability
(\pnull) that a correlation is not present.

A striking regularity emerges from the correlation matrix that
\emph{the velocity shifts of the core and wing components
and the whole profile,
correlate significantly with the EW of the whole \oiii\ emission
($\pnull \lesssim  10^{-5}$) and particularly with that of
the core component, EW(core) ($\pnull \lesssim  10^{-6}$).}
The strongest correlation is between EW(core)
and $\Delta v$(centroid), the velocity shift of the whole profile
($\rs = 0.51$ and $\pnull \ll 10^{-25}$).
For any a specific kind of velocity shift,
the correlation is most significant with EW(core),
yet not very significant (or not significant at all) with the EW(wing)
(see Table~3).
For the EWs of both the core component and the whole emission,
the correlation is more significant with $\Delta v$(centroid)
rather than with $\Delta v$(wing) or $\Delta v$(core),
yet their correlations with $\Delta v$(core) are still very significant
($\pnull \lesssim 10^{-5}$).

In Fig. 2, we show the relationships between
the EW of the core component and the velocity shifts of core component
and the whole profile.
Note that our $\Delta v$(core) is effectively the same as
the velocity shift measured from the peak of the whole \oiii\ profile
(\,$\Delta v$(peak)\,), because by fitting the core component
one is always getting the peak of the whole \oiii\ profile
as the central wavelength of the 1-Gaussian model.
$\Delta v$(peak) is commonly used in the definition of ``blue outliers''
in the literature (e.g., Zomanov et al. 2002, Boroson 2005, Komossa et al. 2008).

To illustrate the \oiii\ velocity shift--EW relationship,
we divide the entire sample into three subsamples according to
$\Delta v$(core) [namely, $\Delta v$(peak)]
with $\Delta v$(core) $< -150$ \kms,
$-100 \leqslant \Delta v$(core) $< -50$ \kms,
and $\Delta v$(core) $\geqslant -50$ \kms.  We then
construct a composite spectrum for each subsample.
We first subtract broad \hb\ and \feii\ emission from the SDSS spectra,
normalize every spectrum by the average flux density in the continuum window
around 4200~\AA, and then construct the composite in the same manner
as Richards et al. (2002) (see their Fig.~4).
The composite spectra are demonstrated in Fig. 3 (panel a).

The variations in EWs and velocity shifts among the composite spectra
can readily be seen to be
consistent with our correlation analysis in the sense that
\emph{the more the peak of the line is blueshifted,
the more the EWs of the total emission and, particularly,
the core component, decrease dramatically, while the blue wing changes much less}.
This is similar to the case of \civ\ as Richards et al. (2002) found
(see also Richards et al. 2010, Wang et al. 2011).
In addition, there seems to be a trend of a positive connection between
the blueshift of the peak and the strength of the wing component,
which is not very significant according to the above correlation result
between $\Delta v$(core) and EW(wing) yet ($\pnull = 4$\%).

\subsection{Correlations with physical AGN parameters}

It is interesting to explore whether the blueshifting/Baldwin effect
of \oiii\,$\lambda5007$ are driven by some fundamental physical parameter.
We have therefore investigated the correlations of the velocity shifts and
EWs of the core and wing components and of the total \oiii\ emission
with broad \hb\ FWHM, continuum luminosity $L_{5100}~\equiv~$\lfive,
\mbh, and \lratio, for the sources in the basic sample.
We calculate the black hole masses based on the FWHM of broad \hb\ using
the formalism presented in Wang et al. (2009, their Eqn.~11).
This formalism was calibrated with recently updated reverberation
mapping-based masses and assuming BLR radius $R \propto L^{0.5}$
(Bentz et al. 2009).  The Eddington ratios are calculated assuming
a bolometric luminosity correction $\lbol \approx 9$\,\lfive\
(Elvis et al. 1994; Kaspi et al. 2000).
The correlation results are summarized in Table~4.

We find that there are some significant correlations ($\pnull \lesssim 10^{-5}$).
The EW of the core component has a negative correlation with
the continuum luminosity (i.e., the classical Baldwin effect) and
with the Eddington ratio (see also Dong et al. 2011).  There is also a
positive correlation of the blueshifting with the Eddington ratio, but none of these
correlations are strong ($\rs < 0.4$).  In particular,
{\em no correlation of any EW or velocity shift with
these physical AGN parameters is more significant (stronger) than
the $\Delta v$(centroid)--EW(core) relationship for the same sample.}

We also divide the 383 sources into three subsamples
by $L_{5100}$ and three subsamples by \lratio.
We construct composite spectra for each subsample as described in Section~3.1.
The three $L_{5100}$ subsamples have
$L_{5100} \geq  10^{44.3}$, $10^{44.3} > L_{5100} \geq 10^{43.8}$,
and $L_{5100} < 10^{43.8}$ \lum, respectively, and
the three \lratio\ subsamples have
$\lratio \geq  10^{-0.4}$, $10^{-0.4} > \lratio \geq 10^{-0.8}$,
and $\lratio < 10^{-0.8}$.
The composite spectra are shown in panels b and c of Fig. 3.
All of the already mentioned correlations can be seen.
We also plot the relatively strong correlations of the velocity shifts
of the peak (i.e., the core) and centroid of \oiii\,$\lambda 5007$
with Eddington ratio (Fig. 4).

In summary, the results worthy of note in the correlation matrix are as follows:
\begin{itemize}
    \item
    The most significant correlation in the EW-related correlation matrix is
    the positive one between the intensity ratio of the wing to the core component
    and the continuum luminosity ($\pnull = 8 \times 10^{-10}$ for the 383 sources),
    which hints a positive correlation---albeit a weak one to be sure---between
    the EW of the wing component and the continuum luminosity.
    This is confirmed by our analysis using the extended sample free from the
    redshift cutoff required by the presence of \sii\ (see below),
    which gives a significant correlation between
    EW(wing) and $L_{5100}$ with $\pnull = 6 \times 10^{-18}$.  The significance is high
    because of the large number of sources (1951) but is still not strong ($\rs \approx 0.2$).
    The correlation of the EW of the core component with luminosity
    is slightly weaker than that with Eddington ratio in the basic sample
    (and much weaker in the extended sample --- see below).  Both are
    much more significant than those with broad-line FWHM or black hole mass.

    \item
    The most significant correlations of all the velocity shifts of the core and wing components
    and of the total profile are with \lratio\ ($\pnull \leq 10^{-9}$),
    rather than with broad-line FWHM, $L_{5100}$ or \mbh.
    These correlations are in the sense that
    the higher the \lratio\, the more the line is blueshifted.
    Moreover, all the three kinds of velocity shifts
    have \emph{no} significant correlation with \mbh\ at all ($\pnull > 10\%$).
    Furthermore, all the objects with large blueshifts have a high Eddington ratio,
    but the reverse is not true (see Fig.~4; also Komossa et al. 2008).
\end{itemize}

The dynamic ranges of $L_{5100}$, \mbh\ and \lratio\ in the basic sample
are 2, 1.5 and 1.5~dex respectively, which are not very large.
It is thus possible that some weak but real correlations are obscured
by the significant measurement errors (see Section~4.1 below;
also Wang et al. 2009, Dong et al. 2011).
To check the correlations concerning EWs that do not require \sii\
as the reference for systematic redshift,
we use the extended sample as defined in Section~2.1.
This sample enlarges the dynamic ranges
of $L_{5100}$, \mbh\ and \lratio\ by $\sim 1$, 1, and 0.5~dex respectively.
The results are also summarized in Table~4 (bottom panel).
The main conclusions remains almost unchanged with similar correlation strengths,
except for the EW of the wing component. As above mentioned, EW(wing) now has
a very significant (yet still not strong) correlation with $L_{5100}$.
It has a less significant correlation with \mbh, and almost no correlation
with \lratio\ still.
Another notable change is that the significance of the correlation of
EW(core) with \lratio\ now becomes much greater than that with $L_{5100}$,
although the correlation is still not tight.

To check for possible effects of the formalism we use to estimated the
black hole masses, we re-examine the correlation
analysis with \mbh\ calculated using several other commonly used formalisms
(e.g., Greene \& Ho 2005, Collin et al. 2006,
Vestergaard \& Peterson 2006, Shen et al. 2010).
All the tests give similar results to those listed in Table~4
(see also Table~3 of Dong et al. 2011).
This is mainly because the dynamic range of \mbh\ covered
by our samples is not very large
($\sim 2.5$ dex for the extended sample, centered at $\sim 10^{8}$ \msun),
and in this range the various formalisms for calculating the black hole masses
have only subtle differences from one another (Wang et al. 2009).

Furthermore, we consider possible effect that our sample selection
and spectra analysis may bring. In Section 2.1, we simply use
the limiting radio flux of FIRST to minimize the effect of possible AGN jets.
It is possible that many of the excluded radio sources are not radio loud
but just nearby and bright.
To check for this, we perform the correlation tests using the 565 sources
before the radio exclusion.
It turns out that, the inclusion of the radio sources
only increases the correlation coefficients by a few per cent (rarely by 10 per cent).
Particularly, all the correlations we concern in this paper,
such as the blueshift--EW(core) relationship, stay nearly unchanged.
This is mainly because the radio sources account for only 30 per cent
of the 565 sources.

We check the effect of using narrow \hb\ as the fiducial reference
for the systematic redshift.
When it is applied to the basic sample, the strengths of the correlations
concerning \oiii\ velocity shifts as listed in Tables~3 and 4
are all reduced significantly but the relative strengths among the correlations
keep almost unchanged. When it is applied to the extended sample,
the existing correlations get more statistically significant due to
the larger sample size, yet the strengths (\rs) get weaker; e.g.,
$\rs = 0.34$ ($\pnull \ll 10^{-100}$) for the correlation of $\Delta v$(centroid)
with EW(core), and $\rs = 0.29$ ($\pnull = 10^{-41}$) for its correlation
with \lratio. Again, the relative strengths among the correlations
keep almost unchanged.

It is, obviously, rather artificial to decompose the \oiii\ profile
into ``core'' and ``wing'' components by using simply two Gaussians.
There is, for example, no theoretical justification that the wing component
should have a Gaussian profile.
And, decomposing the line profile into different components certainly depends
on the line width at the SDSS spectral resolution.
For objects with small line widths and small velocity shifts,
the decomposition would be more difficult.
We do several investigations to check possible decomposition effects.
We find that the width of the decomposed core component,
which is the most sensitive one to the spectral resolution, has only
weak ($\rs< 0.1$) correlations with EW(wing), EW(total), and the velocity shifts
of the core and wing components and of the total emission, and no correlation
($\pnull = 30$\%) with EW(core) at all.
The width of the decomposed wing component has no strong correlations with
the EWs or velocity shifts of the core component
and of the total emission too, and almost no correlations with
either the EW or velocity shift ($\pnull > 1$\%) of the wing component.
These mean that the decomposition process would not bring a serious systematic
effect into our results.

\section{Conclusion and Discussions}

\subsection{\oiii\ properties are not governed preponderantly by AGN parameters?}

Despite the crude nature of the decomposition,
the primary results about the velocity shifts and the EWs of \oiii\,$\lambda5007$
(and its core and wing components)
appear quite robust, according to both the correlation analysis and
the composite spectrum analysis.
Our main result is that, while there is a fairly strong correlation between the EW of
the core component and the velocity shifts, there are no similarly strong correlations of
the EWs or velocity shifts of either the cores, wings, or of the whole emission
with black hole mass, Eddington ratio, or AGN luminosity.
Thus a tentative conclusion is that neither the large object-to-object variation of
the EW of \oiii\,$\lambda 5007$ nor its blueshift are readily attributable to some
fundamental AGN parameters such as $L_{5100}$, \mbh\ and \lratio\
(see also Baskin \& Laor 2005b).
Likewise, neither does the velocity shift--EW
relationship seem to be primarily driven by some AGN parameter.
The weakness of the correlations of \oiii\ properties with fundamental
AGN parameters (including nuclear luminosity, \mbh\ and \lratio)
have also been reported in the literature by using samples with much
larger dynamic ranges (e.g., Dietrich et al. 2002, Netzer et al. 2004).

This is in sharp contrast with the results for the EWs of broad
\mgii\,$\lambda 2800$ doublet emission lines and of narrow optical \feii\ emission
lines, and the intensity ratios of ultraviolet and optical \feii\ to \mgii,
as recently discovered by Dong et al. (2009a, 2011).
They found that those EWs and intensity ratios correlate with \lratio\ strongly
($\rs \gtrsim 0.5$ and $\pnull \ll 10^{-15}$; see the above references),
by using the same data set as the present study.
Considering the effects of measurement errors are similar on the correlations
of both those lines and \oiii, we believe that
the correlations of \oiii\ properties with $L_{5100}$, \mbh, and \lratio\
are intrinsically weak.
For those emission lines arising from the BLR and inner NLR
(narrow \feii\ lines, Dong et al. 2010),
Dong et al. (2009a, 2009b, 2011) proposed that the essential physical mechanism
is that Eddington ratio regulates the global properties (particularly,
the distribution of column density) of the clouds gravitationally bound
in the line-emitting regions.
It appears now that the mechanism operating in the BLR and the inner NLR
does not play a major role in the canonical NLR located far out
in the host galaxies and free from the
gravitational influence of the central supermassive black holes.
{\em A useful positive inference is that,
at least in a statistical sense, the above-mentioned use of
\oiii\,$\lambda5007$ as the proxy of AGN luminosity does not depend
seriously on $L_{5100}$, \mbh\ and \lratio.}

On the other hand, although the \oiii\ properties do not depend as strongly on
$L_{5100}$, \mbh\ or \lratio\ as the above-mentioned emission lines
arising from the inner NLR and the BLR, there
are still some significant (albeit not strong) correlations
between \oiii\ properties and those fundamental AGN parameters.
The EW of the core component has the most significant (anti-)correlation with
\lratio, a less significant correlation with continuum luminosity,
a much lower correlation with the broad-line FWHM, and almost no correlation with \mbh.
For the wing component of \oiii\ the most significant correlation
is with continuum luminosity.
The behavior of the total emission results from the combination of
the above two factors, and is mainly dominated by the core component.
The magnitude of the blueshifting of the core, the wing,
and the whole profile has the most significant correlation with \lratio,
rather than with the other parameters. Because the SDSS spectroscopic survey is
magnitude-limited and, moreover, because both \mbh\ and \lratio\
are constructed from broad-\hb\ FWHM and $L_{5100}$, there are apparent
(likely not intrinsic) correlations among these four quantities.
For instance, the Spearman correlation coefficients of \lratio\
with broad-\hb\ FWHM, $L_{5100}$, and \mbh\ for our basic sample of 383 sources
are \rs\ = $-0.82$, $0.30$, and $-0.51$, respectively.
In light of the serious inter-dependence among these four quantities,
the correlations of EW(wing) with broad-\hb\ FWHM, \mbh, or \lratio\
are probably a secondary effect of the stronger
(thus presumably intrinsic) correlation with $L_{5100}$, while
the correlations of the velocity shifts and EW(core) with
broad-\hb\ FWHM, $L_{5100}$, or \mbh\ are a secondary effect
of that with \lratio. This is probable given that there are additional systematic
uncertainties plaguing the estimated values of \lratio.
One effect is the large uncertainties in virial BH masses, which can
be a factor of 4 ($1\sigma$) statistically, and perhaps as large as an order
of magnitude for individual estimates (Vestergaard \& Peterson 2006; Wang et
al. 2009). Another uncertainty comes from the bolometric correction assumed
for $L_{5100}$, which is definitely an oversimplification given the diverse
spectral energy distributions of AGNs (Ho 2008; Vasudevan \& Fabian 2009;
Grupe et al. 2010).
Yet, since the above bivariate correlations of \oiii\ properties
with $L_{5100}$, \mbh\ and \lratio\ are not strong, being weaker than
the $\Delta v$(centroid)--EW(core) relationship and
much weaker even than the inter-dependence among the AGN parameters,
we refrain from an over-discussion on this issue at present.
The weakness of those correlations either reflects true scatters that are
caused by multiple (and somehow independent) processes and are thus irreducible,
or suggests that there are other dominant variable(s) yet to be identified
(see the discussions on various possible physical variables/processes
in Baskin \& Laor 2005b, Risaliti et al. 2011 and
Caccianiga \& Severgnini 2011).

\subsection{The velocity shift--EW(core) connection}

The fairly tight anticorrelation between the blueshifts and the EW of the core
component appears more notable than the weak correlations with fundamental AGN
properties.  Interestingly, as analyzed in Subsection 3.1, this pattern is just
like the situation for the broad \civ\,$\lambda1549$ emission line
(Richards et al. 2002, Bachev et al. 2004, Richards et al. 2010),
which has already been noted by some researchers (e.g., Zamanov et al. 2002).
This suggests that both the blueshifting
phenomenon and the classical Baldwin effect are tightly linked (Corbin 1991, 1992).
Note that the classical Baldwin effect only concerns the core component.  The wing
component can even show an inverse Baldwin effect---the higher the AGN luminosity,
the larger the EW of the wing component.  This is similar to the situation of
\civ\,$\lambda1549$ (Richards et al. 2010, Wang et al. 2011).
%This supports the proposal of Brotherton et al. (1994) that the low-velocity BLR gas
%(the gas responsible for the line cores) is intermediate between the BLR and the NLR.

In light of the remarkable similarity in the velocity shift--EW connection
between \oiii\ and \civ, Occam's razor would suggest that
the mechanism producing the two correlations is the same.
A wind has been proposed as the cause of the \civ\ blueshifting (Gaskell 1982)
and the correlations discussed above seem to be consistent with a scenario where
the core of \oiii\ comes from the canonical extended NLR where the gas motions
are dominated by the gravity of the bulge stars while the wing component
arises from an outflow.  This has been discussed by many authors
(see Zamanov et al. 2002, Baskin \& Laor 2005a, Richards et al. 2010,
and Wang et al. 2011 for details).
In this model the ``blue outliers'' could just be extremes;
presumably a higher relative accretion rate drives a stronger outflow.

As pointed out in Section 4.1, the large object-to-object variation in
the strength of \oiii\ emission
and the blueshift--EW(core) connection are not governed primarily by any
fundamental AGN parameters such as $L_{5100}$, \mbh\ and \lratio.
It is likely that it is the ISM conditions of the host galaxies that mainly
determine the diversity of the \oiii\ properties in active galaxies
(particularly its strength; see Netzer et al. 2004).
In the outflow scenario, it is natural to assume that the ISM gives
rise to \oiii\ emission and decelerates the outflow.
Although AGN activity
determines the launching speed of outflow, the final speed of the
outflow depends on the density and density profile of the ISM.
Meanwhile, the density and the density profile also
constrain the amount of \oiii-emitting gas and its emissivity.
At the same AGN activity level, denser ISM will
produce more \oiii\ emission while decelerate
the outflow efficiently, leading to higher EW and lower velocity.
On the other hand, the similar blueshift--EW relationship in \civ\ may be
linked to the shape of the ionizing continuum (Richards et al. 2010).
So it is possible that the above two factors work together
to produce the blueshift--EW(core) connection.

Yet, we should also mention that,
while the outflow scenario is attractive, there seem to be problems.
For instance, the outflow explanation for the blueshifting of \civ\
(Gaskell 1982) clashes head on with the BLR kinematics
implied by velocity-resolved reverberation mapping.
Velocity-resolved kinematics have long favored gravitationally-dominated motions
(Gaskell 1988, Koratkar \& Gaskell 1989, 1991; Gaskell 2009b, 2010b) and apparent
cases of outflow are unlikely to be real but are instead probably the result of
off-axis variability (Gaskell 2010a).  Gaskell \& Goosmann (2008) have proposed
instead that the blueshifting is the result of scattering off material with a
slight net inflow.  This naturally explains the correlation of blueshifting with
accretion rate.  If the NLR and BLR blueshiftings have a similar cause,
the blueshifting of \oiii\ too might be caused by scattering,
as has been proposed by Gaskell \& Goosmann (2008).

Finally, we note that while all the AGNs with large blueshifts
have a high Eddington ratio, the reverse is not true (see also Komossa et al. 2008).
Again, this is very similar to the situation for broad \civ\,$\lambda 1549$
(Baskin \& Laor 2005a).  Here we must stress once again (see Section 4.1) that
the correlations between velocity shift and any fundamental
AGN parameter such as nuclear luminosity, black hole mass, and the Eddington ratio, are
not strong (see also Boroson 2005, Komossa et al. 2008). Thus it is not clear
if the velocity shift--EW(core) connection for \oiii\,$\lambda5007$ can be
reduced to the effect of a specific underlying physical process of the AGN activity.
Anyway, whatever the cause(s) of the blueshifting of high-ionization NLR and BLR lines,
the correlations discussed here provide further support
for an intimate connection between the NLR and the BLR.

\acknowledgements
We thank the anonymous referee for helpful suggestions,
Jian-Guo Wang and Dawei Xu for discussions, and
Yifei Chen for computing support.
This work is supported by Chinese NSF grants NSF-10703006, NSF-10973013, and NSF-11073019,
a National 973 Project of China (2009CB824800), %---xbdong
and the Fundamental Research Funds for the Central Universities (USTC WK2030220004).
The visit of X.-B.\,D. to Carnegie Observatories is mainly supported by OATF, USTC
(up to April 2011). MG is grateful for support from the
GEMINI-CONICYT Fund through grant 32070017 and
from the Cento de Astrof\'isica de Valpara\'iso.
Funding for the SDSS and SDSS-II has been provided by the Alfred P.
Sloan Foundation, the Participating Institutions, the National
Science Foundation, the U.S. Department of Energy, the National
Aeronautics and Space Administration, the Japanese Monbukagakusho,
the Max Planck Society, and the Higher Education Funding Council for
England.  The SDSS Web Site is http://www.sdss.org/.
\newline

%%%%%%%%%%%%%%%%%%%%%%%%%%%%%%%%%%%%%%%%% figures
\begin{figure}
\begin{center}
\includegraphics{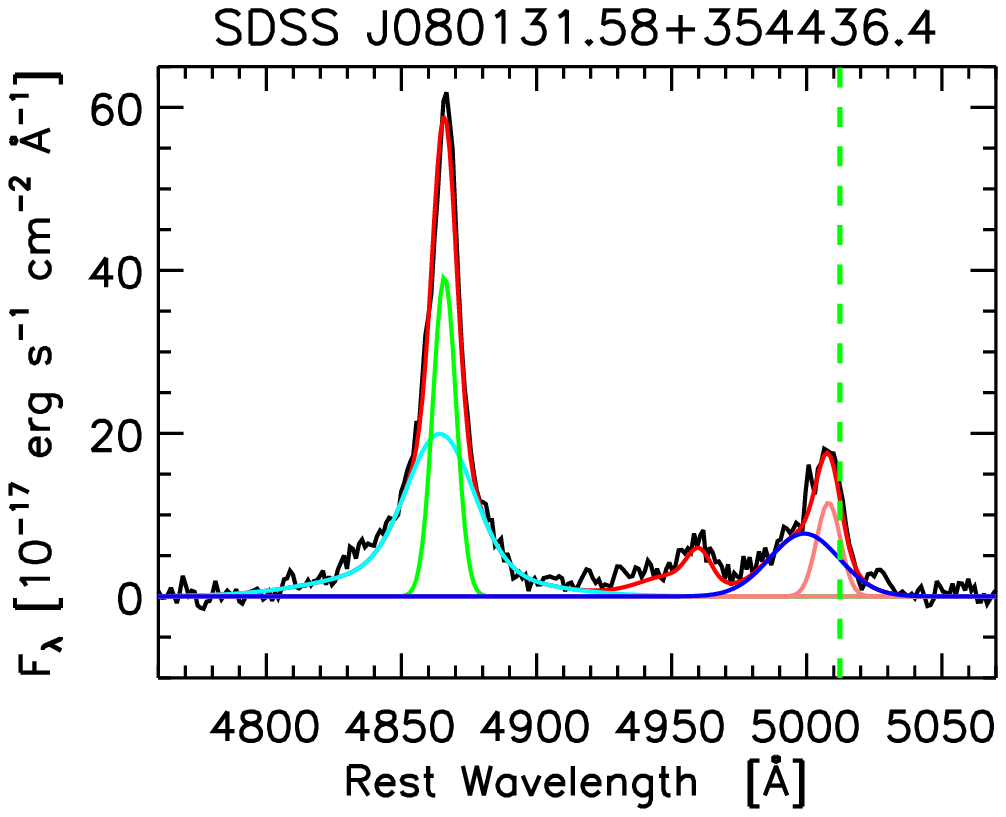}
\caption{
Results of the emission-line profile fitting applied to the \hb--\oiii\ region
of SDSS J080131.58+354436.4.
We plot the continuum and \feii\ subtracted spectrum (black),
the sum of all the best-fitting components (red),
the fitted narrow (green) and broad \hb\ (cyan),
and particularly the decomposed core (pink) and wing (blue) components
of \oiii\,$\lambda5007$ line.
The green dashed vertical line denotes the wavelength for \oiii\,$\lambda5007$
as inferred from the \sii\ line. \label{fig:fig-1}}
\end{center}

\end{figure}

%%%%%%%%%%%%%%%%%% Figure 2 %%%%%%%%%%%%%%%%%%%%%%%%%%%%%%%
\begin{figure}
\begin{center}
\includegraphics{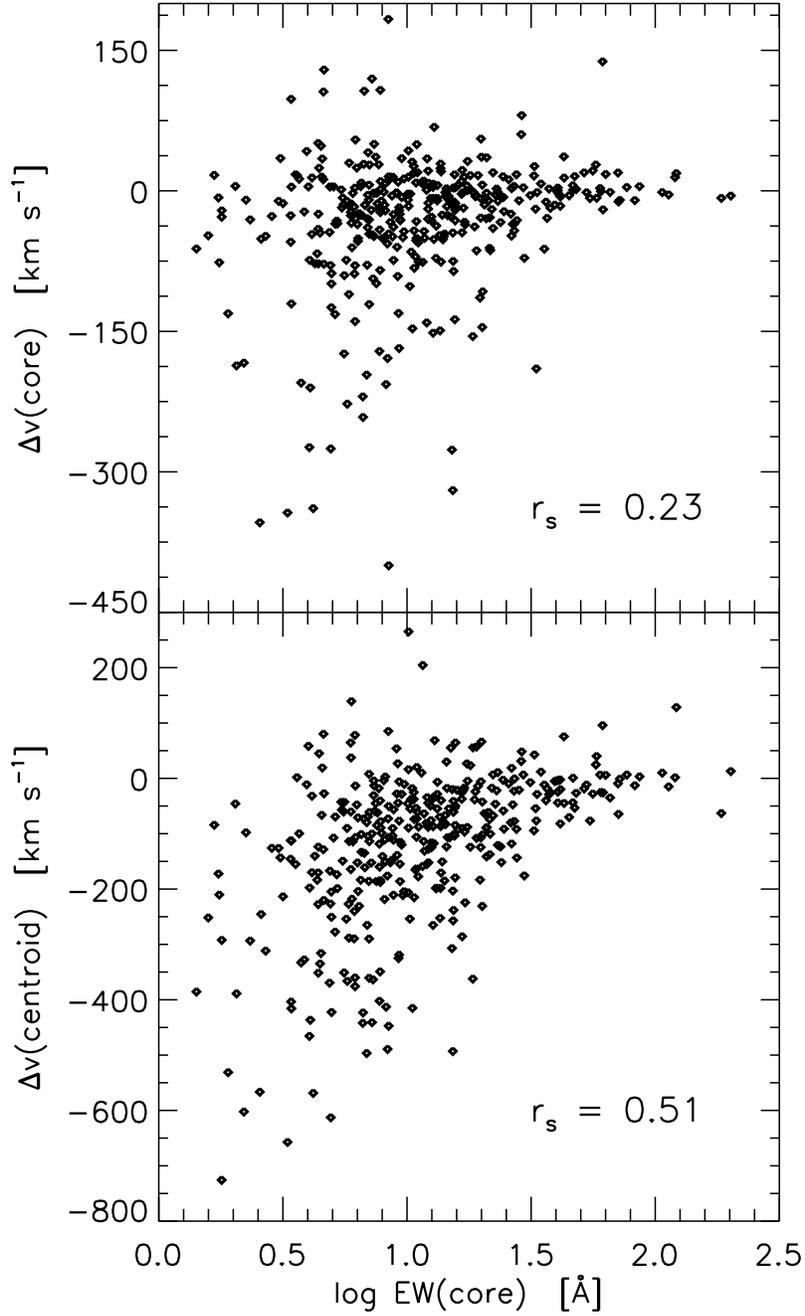}
\caption{
Distributions of velocity shifts of the core component (top) and
the centroid of the whole profile (bottom) of \oiii\,$\lambda5007$
versus the equivalent width of the core component, for the basic sample
of 383 objects.
\label{fig:fig-ew-vshift}
}
\end{center}
\end{figure}
%\clearpage

%%%%%%%%%%%%%%%%%% Figure 3 %%%%%%%%%%%%%%%%%%%%%%%%%%%%%%%
\begin{figure}
\begin{center}
\includegraphics{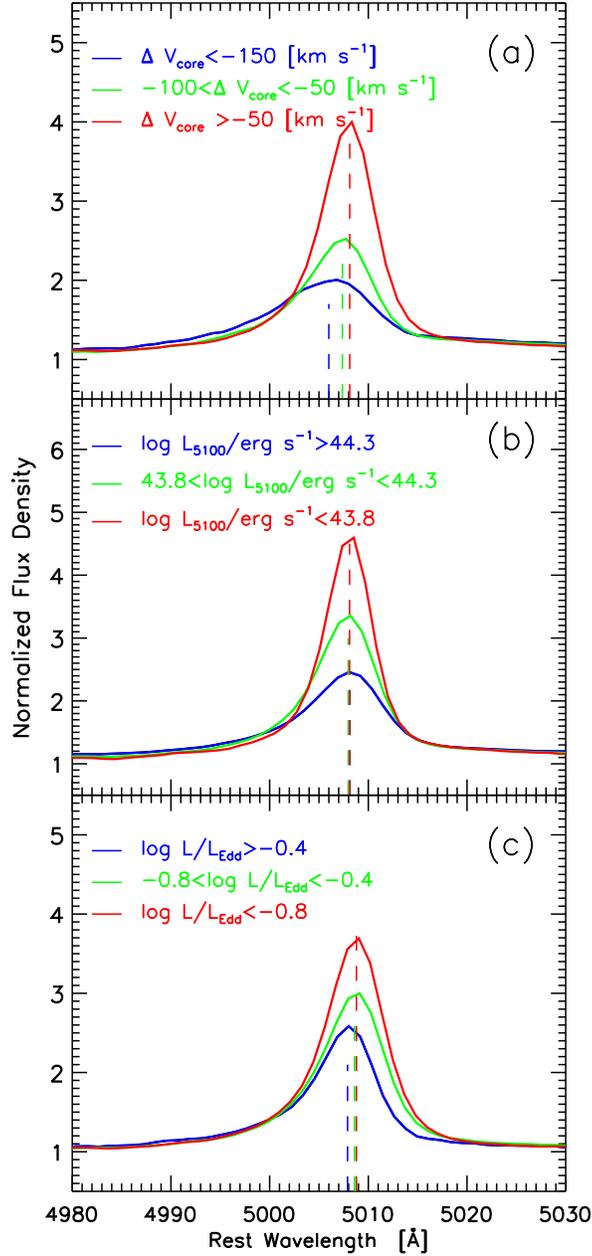}
\caption{Panel (a): The composite spectrum of
different $\Delta v$(core) bins. The blue, green and red lines are
spectra with decreasing $\Delta v$(core).
Panel (b): The composite spectrum of different $L_{5100}$
bins. The blue, green and red lines are spectra with decreasing
$L_{5100}$. Panel (c): The composite spectrum of different
$L/L_{Edd}$ bins. The blue, green and red lines are spectra with
decreasing $L/L_{Edd}$.
\label{fig:fig-composite}}
\end{center}
\end{figure}
%\clearpage

%%%%%%%%%%%%%%%%%% Figure 4 %%%%%%%%%%%%%%%%%%%%%%%%%%%%%%%
\begin{figure}
\begin{center}
\includegraphics{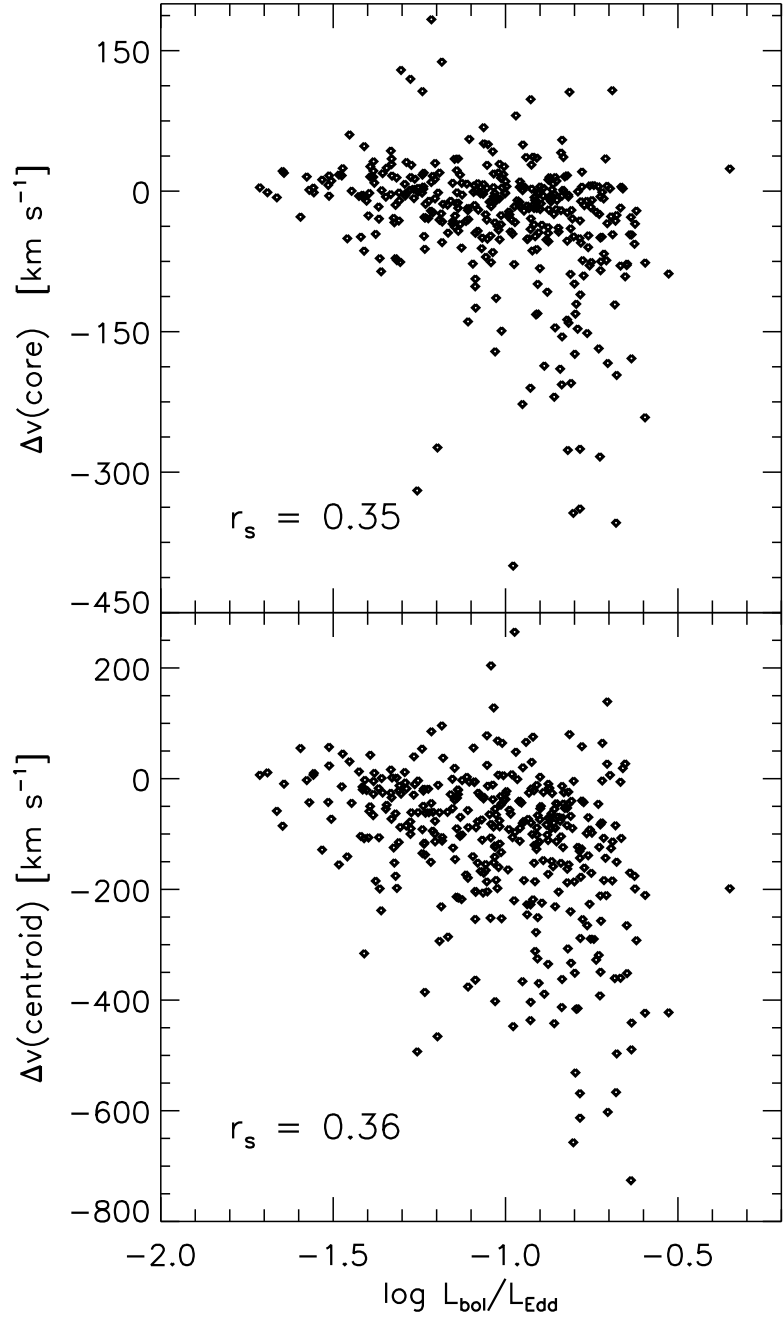}
\caption{
Distributions of velocity shifts of the core component (top) and
the centroid of the whole profile (bottom) of \oiii\,$\lambda5007$
versus the Eddington ratio, for the basic sample
of 383 objects.
\label{fig:fig-vshift-lratio}
}
\end{center}

\end{figure}

\clearpage

%%%%%%%%%%%%%%% Table 1 %%%%%%%%%%%%%%%%%%%%%%%%%%%%%%%
\begin{deluxetable}{lccccccc}
%\centering
\tabletypesize{\scriptsize}
%\topmargin 0.0cm \evensidemargin = 0mm
%\oddsidemargin=0mm
%\rotate
\tablecaption{Parameters of the Core and Wing Components of
\oiii\,$\lambda 5007$ Emission Line:
Basic Sample}
\tablehead{
\colhead{SDSS Name} &
\colhead{$\log F$(core)} &
\colhead{EW(core)} &
\colhead{$\log F$(wing)} &
\colhead{EW(wing)} &
\colhead{$\Delta v$(core)} &
\colhead{$\Delta v$(wing)} &
\colhead{$\Delta v$(centroid)} \\
\colhead{(1)}  & \colhead{(2)} & \colhead{(3)} & \colhead{(4)} & \colhead{(5)} &
\colhead{(6)}  & \colhead{(7)} & \colhead{(8)}
}
\startdata
J000410.80-104527.2  &    -15.02  &     4.39   &   -15.00   &    4.64  &    -77.52   &   -608.9  &    -351.3  \\
J000904.54-103428.6  &    -14.24  &    33.43   &   -14.59   &   14.93  &    -1.431   &    -88.4  &     -28.3  \\
J001247.93-084700.4  &    -14.75  &     6.55   &   -14.46   &   12.91  &    -25.71   &   -186.9  &    -132.7  \\
J001340.73-111100.6  &    -14.56  &    15.29   &   -15.13   &    4.10  &    -319.6   &  -1134.4  &    -492.9  \\
J001416.92+145038.4  &    -14.94  &     9.24   &   -14.85   &   11.26  &    -130.5   &   -483.8  &    -324.9  \\
J003723.49+000812.5  &    -14.82  &    13.12   &   -14.97   &    9.35  &    -37.27   &   -422.4  &    -197.9  \\
J004930.90+153216.3  &    -14.08  &    32.48   &   -14.62   &    9.24  &     16.22   &   -480.5  &     -94.2  \\
J005118.27+135448.0  &    -14.07  &    26.38   &   -15.26   &    1.72  &     2.772   &    265.9  &      18.8  \\
J005328.80-085755.0  &    -14.69  &    10.91   &   -14.98   &    5.51  &    -32.51   &   -250.1  &    -105.7  \\
J005709.93+144610.2  &    -14.02  &     6.72   &   -14.19   &    4.57  &     106.7   &   -487.3  &    -134.7
\enddata
\tablecomments{\normalsize Parameters for the 383 objects in the basic sample.
Column (1) official SDSS name;
Column (2) Flux of the core component;
Column (3) Equivalent Width of the core component;
Column (4) Flux of the wing component;
Column (5) Equivalent width of the wing component;
Column (6) Velocity shift of the core component,
with respect to \sii\,$\lambda\lambda 6716,6731$ doublet lines;
Column (7) Velocity shift of the wing component, with respect to \sii;
Column (8) Velocity shift of the centroid, with respect to \sii.
(This table is available in its entirety in a machine-readable form in the online
journal. A portion is shown here for guidance regarding its form and content.)}
\end{deluxetable}

%%%%%%%%%%%%%%%% end of Table 1 %%%%%%%%%%%%%%%

%%%%%%%%%%%%%%% Table 2 %%%%%%%%%%%%%%%%%%%%%%%%%%%%%%%
\begin{deluxetable}{lccccccc}
%\centering
\tabletypesize{\scriptsize}
%\topmargin 0.0cm \evensidemargin = 0mm
%\oddsidemargin=0mm
%\rotate
\tablecaption{EW and Flux of the Core and Wing Components of
\oiii\,$\lambda 5007$ Emission Line:
Extended Sample}
\tablehead{
\colhead{SDSS Name} &
\colhead{$\log F$(core)} &
\colhead{EW(core)} &
\colhead{$\log F$(wing)} &
\colhead{EW(wing)} \\
\colhead{(1)}  & \colhead{(2)} & \colhead{(3)} & \colhead{(4)} & \colhead{(5)}
}
\startdata
J000043.95-091134.9   &   -15.27   &   4.38   &   -15.42  &    3.11    \\
J000102.19-102326.8   &   -14.39   &   35.43  &   -14.66  &    18.9    \\
J000110.96-105247.4   &   -14.99   &   5.52   &   -14.70  &    10.8    \\
J000545.61+153833.8   &   -14.88   &   7.73   &   -14.59  &    15.0    \\
J000945.46+001337.1   &   -15.73   &   2.77   &   -15.03  &    13.7    \\
J001030.58+010006.0   &   -13.84   &   82.26  &   -14.43  &    21.4    \\
J001327.31+005232.0   &   -14.33   &   24.42  &   -14.34  &    23.6    \\
J001630.43-093853.4   &   -14.98   &   7.48   &   -15.00  &    7.18    \\
J001725.35+141132.5   &   -14.74   &   5.28   &   -14.52  &    8.82    \\
J001855.22-091351.1   &   -15.08   &   3.85   &   -14.70  &    9.39
\enddata
\tablecomments{\normalsize Parameters for the 1568 objects that are not in the basic sample
but in the extended sample.
Column (1) official SDSS name;
Column (2) Flux of the core component;
Column (3) Equivalent Width of the core component;
Column (4) Flux of the wing component;
Column (5) Equivalent width of the wing component.
 (This table is available in its entirety in a machine-readable form in the online
journal. A portion is shown here for guidance regarding its form and content.)}
\end{deluxetable}
%%%%%%%%%%%%%%%%%%% End of Table 2 %%%%%%%%%%%%%%%%%%%%%%%%%%%%%%

%%%%%%%%%%%%%%%%% Table 3 %%%%%%%%%%%%%%%%%%%%%%%%%%%%%%%%%%%%%%%
\begin{table*}[h]
%\centering
\topmargin 0.0cm
\evensidemargin = 0mm
\oddsidemargin = 0mm
\small %\tiny %
\caption{Correlation Results between Velocity Shift and  Equivalent Width \tablenotemark{a}}
\label{corrtab}
\medskip
\vfill
\begin{tabular}{l|c c c c}
\hline \hline
                     &  EW(core) & EW(wing) & EW(total)\\
\hline
$\Delta v$(core)     & 0.23 (4e-06) & 0.10 (4e-02) & 0.20 (7e-05)  \\
$\Delta v$(wing)     & 0.41 (6e-17) & 0.19 (2e-04) & 0.38 (1e-14) \\
$\Delta v$(centroid) & 0.51 (2e-26) & 0.11 (3e-02) & 0.41 (8e-17) \\
\hline
\end{tabular}
\medskip
\vfill
{\normalsize $^a$~For each entry, we list the Spearman rank correlation
      coefficient (\rs) and the probability of the null hypothesis that
      the correlation is not present (\pnull) in parenthesis,
      for the 383 objects in the basic sample.}\\
%\end{sidewaystable*}
\end{table*}
%%%%%%%%%%%%%%%% end of Table 3 %%%%%%%%%%%%%%%

%%%%%%%%%%%%%%%%% Table 4 %%%%%%%%%%%%%%%%%%%%%%%%%%%%%%%%%%%%%%%
\begin{table*}[h]
%\centering
\topmargin 0.0cm
\evensidemargin = 0mm
\oddsidemargin = 0mm
\small %\tiny %
\caption{Correlation Results between \oiii\ Quantities and
Physical Parameters \tablenotemark{a}}
\label{corrtab}
\medskip
\vfill
\begin{tabular}{l|c c c c c}
\hline \hline
                                   & N            \tablenotemark{b}
& FWHM(H$\beta^{\rm B}$)
& $L_{5100}$   \tablenotemark{c}
& \mbh         \tablenotemark{c}
& \lratio      \tablenotemark{c}                 \\
\hline
$\Delta v$(core)                   &  383 &    0.28 (3e-08) &      -0.12 (1e-02)  &     0.17 (7e-04) &        -0.35 (3e-12)        \\
$\Delta v$(wing)                   &  383 &    0.17 (7e-04) &      -0.25 (1e-06)  &     0.03 (5e-01) &        -0.31 (1e-09)      \\
$\Delta v$(centroid)               &  383 &    0.22 (2e-05) &      -0.27 (6e-08)  &     0.06 (2e-01) &        -0.36 (2e-13)      \\
EW(core)                           &  383 &    0.14 (5e-03) &      -0.26 (2e-07)  &    -0.01 (9e-01) &        -0.28 (4e-08)   \\
EW(wing)                           &  383 &    0.10 (6e-02) &       0.02 (7e-01)  &     0.08 (1e-01) &        -0.07 (1e-01)       \\
EW(total)                          &  383 &    0.14 (5e-03) &      -0.18 (3e-04)  &     0.03 (6e-01) &        -0.23 (5e-06)       \\
$\frac{\rm EW(wing)}{\rm EW(core)}$&  383 &   -0.06 (3e-01) &       0.31 (8e-10)  &     0.09 (8e-02) &         0.22 (1e-05)        \\
\hline
EW(core)                           & 1951 &    0.11 (1e-06) &      -0.16 (3e-12)  &    -0.02 (4e-01) &        -0.22 (6e-23)        \\
EW(wing)                           & 1951 &    0.08 (5e-04) &       0.19 (6e-18)  &     0.16 (9e-13) &         0.07 (1e-03)       \\
EW(total)                          & 1951 &    0.12 (7e-08) &       0.01 (6e-01)  &     0.08 (3e-04) &        -0.11 (3e-06)      \\
$\frac{\rm EW(wing)}{\rm EW(core)}$& 1951 &   -0.05 (2e-02) &       0.31 (6e-44)  &     0.14 (5e-10) &         0.28 (3e-36)        \\
\hline
\end{tabular}
\medskip
\vfill
{\normalsize $^a$~For each entry, we list the Spearman rank correlation
      coefficient (\rs) and the probability of the null hypothesis that
      the correlation is not present (\pnull) in parenthesis.}\\
{\normalsize $^b$~The number of the sources used in every bivariate correlation tests.}\\
{\normalsize $^c$~$L_{5100} \equiv \lambda L_{\lambda}$(5100\,\AA);
the BH masses are calculated using
      the formalism presented in Wang et al. (2009, their Eqn.~11);
      Eddington ratios (\lratio) are calculated assuming
      that the bolometric luminosity $\lbol \approx 9\,L_{5100}$.}
%\end{sidewaystable*}
\end{table*}
%%%%%%%%%%%%%%%% end of Table 4 %%%%%%%%%%%%%%%

%\end{CJK}

\begin{thebibliography}{}

\bibitem[Adelman-McCarthy et al.(2006)]{2006ApJS..162...38A}
Adelman-McCarthy, J.~K., et al.\ 2006, \apjs, 162, 38

\bibitem[Baldwin(1977)]{1977ApJ...214..679B} Baldwin, J.~A.\ 1977, \apj,
214, 679

\bibitem[Baldwin et al.(1978)]{1978Natur.273..431B} Baldwin, J.~A., Burke,
W.~L., Gaskell, C.~M., \& Wampler, E.~J.\ 1978, \nat, 273, 431

%\bibitem[\protect\citeauthoryear{Baldwin}{1977}]{baldwin77} Baldwin, J.~A.\ 1977, ApJ, 214, 679.
%% Luminosity Indicators in the Spectra of Quasi-Stellar Objects

%\bibitem[\protect\citeauthoryear{Baldwin et al.}{1978}]{baldwin+78} Baldwin, J.~A., Burke,
%W.~L., Gaskell, C.~M., \& Wampler, E.~J.\ 1978, Nature, 273, 431.
%% Relative quasar luminosities determined from emission line strengths

\bibitem[Bachev et al.(2004)]{2004ApJ...617..171B} Bachev, R., Marziani,
P., Sulentic, J.~W., Zamanov, R., Calvani, M.,
\& Dultzin-Hacyan, D.\ 2004, \apj, 617, 171

\bibitem[Baskin \& Laor(2005)]{2005MNRAS.356.1029B} Baskin, A., \& Laor, A.\ 2005a, \mnras, 356, 1029

\bibitem[Baskin \& Laor(2005)]{2005MNRAS.358.1043B} Baskin, A., \& Laor, A.\ 2005b, \mnras, 358, 1043

\bibitem[Becker et al.(1995)]{1995ApJ...450..559B} Becker, R.~H., White,
R.~L., \& Helfand, D.~J.\ 1995, \apj, 450, 559

\bibitem[Bentz et al.(2009)]{2009ApJ...705..199B} Bentz, M.~C., et al.\
2009, \apj, 705, 199

\bibitem[Boroson \& Green(1992)]{1992ApJS...80..109B} Boroson, T.~A., \& Green, R.~F.\ 1992, \apjs, 80, 109

\bibitem[Boroson(2005)]{2005AJ....130..381B} Boroson, T.\ 2005, \aj, 130,
381

%\bibitem[\protect\citeauthoryear{Brotherton et al.}{1994}]{brotherton+94} Brotherton, M.~S.,
%Wills, B.~J., Francis, P.~J., \& Steidel, C.~C.\ 1994, \apj, 430, 495
%%  The intermediate line region of QSOs

\bibitem[Caccianiga \& Severgnini(2011)]{2011MNRAS.tmp..758C} Caccianiga, A., \& Severgnini, P.\ 2011, \mnras, 758


\bibitem[Crenshaw et al.(2010)]{2010ApJ...708..419C} Crenshaw, D.~M.,
Schmitt, H.~R., Kraemer, S.~B., Mushotzky, R.~F.,
\& Dunn, J.~P.\ 2010, \apj, 708, 419

\bibitem[Corbin(1990)]{1990ApJ...357..346C} Corbin, M.~R.\ 1990, \apj, 357,
346

\bibitem[Corbin(1991)]{1991ApJ...371L..51C} Corbin, M.~R.\ 1991, \apjl,
371, L51

\bibitem[Corbin(1992)]{1992ApJ...391..577C} Corbin, M.~R.\ 1992, \apj, 391,
577

\bibitem[Collin et al.(2006)]{2006A&A...456...75C} Collin, S., Kawaguchi, T., Peterson,
B.~M., \& Vestergaard, M.\ 2006, \aap, 456, 75

\bibitem[Das et al.(2006)]{2006AJ....132..620D} Das, V., Crenshaw, D.~M.,
Kraemer, S.~B., \& Deo, R.~P.\ 2006, \aj, 132, 620

\bibitem[Dietrich et al.(2002)]{2002ApJ...581..912D} Dietrich, M., Hamann,
F., Shields, J.~C., Constantin, A., Vestergaard, M., Chaffee, F.,
Foltz, C.~B., \& Junkkarinen, V.~T.\ 2002, \apj, 581, 912

\bibitem[Dong et al.(2005)]{2005ApJ...620..629D} Dong, X.-B., Zhou, H.-Y.,
Wang, T.-G., Wang, J.-X., Li, C., \& Zhou, Y.-Y.\ 2005, \apj, 620, 629

\bibitem[Dong et al.(2008)]{2008MNRAS.383..581D} Dong, X., Wang, T., Wang, J.,
Yuan, W., Zhou, H., Dai, H., \& Zhang, K.\ 2008, \mnras, 383, 581

\bibitem[Dong et al.(2009a)]{2009ApJ...703L...1D} Dong, X.-B., Wang, T.-G.,
Wang, J.-G., Fan, X., Wang, H., Zhou, H., \& Yuan, W.\ 2009a, \apjl, 703, L1

\bibitem[Dong et al.(2009b)]{2009ASPC..408...83D} Dong, X.-B., Wang, J.-G.,
Wang, T.-G., Wang, H., Fan, X., Zhou, H., Yuan, W., \& Long, Q.\ 2009b,
in ASP Conf. Ser. 408, The Starburst-AGN Connection, ed. W. Wang et al.
(San Francisco, CA: ASP), 83 (arXiv:0903.0698)

\bibitem[Dong et al.(2010)]{2010ApJ...721L.143D} Dong, X.-B., Ho, L.~C.,
Wang, J.-G., Wang, T.-G., Wang, H., Fan, X.,
\& Zhou, H.\ 2010, \apjl, 721, L143

\bibitem[Dong et al.(2011)]{2009arXiv0903.5020D} Dong, X.-B., Wang, J.-G.,
Ho,~L.~C., Wang,~T., Fan,~X., Wang, H., Zhou, H., \& Yuan, W.\ 2011,
ApJ accepted (arXiv:0903.5020)


\bibitem[Elvis et al.(1994)]{1994ApJS...95....1E} Elvis, M., et al.\ 1994,
\apjs, 95, 1


%\bibitem[\protect\citeauthoryear{Gaskell}{1982}]{gaskell82} Gaskell, C.~M.\ 1982,
%ApJ, 263, 79
%% A redshift difference between high and low ionization emission-line regions in QSOs - Evidence for radial motions

%\bibitem[\protect\citeauthoryear{Gaskell}{1988}]{gaskell88} Gaskell, C.~M.\ 1988,
%ApJ, 325, 114
%% Direct Evidence of Gravitational Domination of the Motion of Gas Within One
%% Light-Week of the Central Object in NGC 4151 and the Determintion of the Mass of the Probable Black Hole

%\bibitem[\protect\citeauthoryear{Gaskell}{2009a}]{gaskell09a} Gaskell, C.~M.\ 2009a,
%submitted to ApJ (arXiv:0908.0328)
%% An Improved [O III] Line Width to Stellar Velocity Dispersion Calibration: Curvature, Scatter, and Lack
%% of Evolution in the Black-Hole Mass Versus Stellar Velocity Dispersion Relationship

%\bibitem[\protect\citeauthoryear{Gaskell}{2009b}]{gaskell09b} Gaskell, C.~M. 2009b,
%New Astron. Rev, 53, 140
%% What broad emission lines tell us about how active galactic nuclei work

%\bibitem[\protect\citeauthoryear{Gaskell}{2010b}]{gaskell10b} Gaskell, C.~M. 2010b,
%in Accretion and Ejection in AGNs: a Global View,
%eds. L. Maraschi, G. Ghisellini, R. Della Ceca \& F. Tavecchio, ASP Conf. Ser, 427, 68
%% Inflow of the Broad-Line Region and the Fundamental Limitations of Reverberation Mapping

%\bibitem[\protect\citeauthoryear{Gaskell \& Goosmann}{2008}]{gaskell+goosmann08}
%Gaskell, C. M. \& Goosmann, R. W. 2008, submitted to ApJ (arXiv:0805.4258)
%% Line Shifts, Broad-Line Region Inflow, and the Feeding of AGNs

\bibitem[Gaskell(1982)]{1982ApJ...263...79G} Gaskell, C.~M.\ 1982, \apj,
263, 79
\bibitem[Gaskell(1988)]{1988ApJ...325..114G} Gaskell, C.~M.\ 1988, \apj,
325, 114

\bibitem[Gaskell
\& Goosmann(2008)]{2008arXiv0805.4258G} Gaskell, C.~M., \& Goosmann, R.~W.\ 2008, arXiv:0805.4258

\bibitem[Gaskell(2009)]{2009NewAR..53..140G} Gaskell, C.~M.\ 2009, New Astronomy Reviews,
53, 140

\bibitem[Gaskell(2009)]{2009arXiv0908.0328G} Gaskell, C.~M.\ 2009,
arXiv:0908.0328

\bibitem[Gaskell(2010)]{2010arXiv1008.1057G} Gaskell, C.~M.\ 2010,
arXiv:1008.1057

\bibitem[Gaskell(2010)]{2010ASPC..427...68G} Gaskell, C.~M.\ 2010,
Accretion and Ejection in AGN: a Global View (San Francisco, CA: ASP), 427, 68


\bibitem[Greene et al.(2009)]{2009ApJ...702..441G} Greene, J.~E., Zakamska,
N.~L., Liu, X., Barth, A.~J., \& Ho, L.~C.\ 2009, \apj, 702, 441

\bibitem[Grupe et al.(2010)]{2010ApJS..187...64G} Grupe, D., Komossa, S.,
Leighly, K.~M., \& Page, K.~L.\ 2010, \apjs, 187, 64

\bibitem[Grindlay et al.(1980)]{1980ApJ...239L..43G} Grindlay, J.~E.,
Forman, W.~R., Steiner, J.~E., Canizares, C.~R.,
\& McClintock, J.~E.\ 1980, \apjl, 239, L43

\bibitem[Heckman et al.(2004)]{2004ApJ...613..109H} Heckman, T.~M.,
Kauffmann, G., Brinchmann, J., Charlot, S., Tremonti, C., \& White,
S.~D.~M.\ 2004, \apj, 613, 109

\bibitem[Ho 2008]{} Ho, L.~C. 2008, ARA\&A, 46, 475

\bibitem[Kaspi et al.(2000)]{2000ApJ...533..631K} Kaspi, S., Smith, P.~S.,
Netzer, H., Maoz, D., Jannuzi, B.~T., \& Giveon, U.\ 2000, \apj, 533, 631

\bibitem[Komossa\& Xu(2007)]{2007ApJ...667L..33K} Komossa, S., \& Xu, D.\ 2007,
\apjl, 667, L33

\bibitem[Komossa et al.(2008)]{2008ApJ...680..926K} Komossa, S., Xu, D.,
Zhou, H., Storchi-Bergmann, T., \& Binette, L.\ 2008, \apj, 680, 926

%\bibitem[\protect\citeauthoryear{Koratkar \& Gaskell}{1989}]{koratkar+gaskell89}
%Koratkar, A. P. \& Gaskell, C. M. 1989, ApJ, 345, 637
%% Emission-Line Variability of Fairall 9: Determination of the Size of the Broad-Line
%% Region and the Direction of Gas Motion.

%\bibitem[\protect\citeauthoryear{Koratkar \& Gaskell}{1991b}]{koratkar+gaskell91b}
%Koratkar, A.~P., \& Gaskell, C.~M.\ 1991b, ApJS, 75, 719
%% Structure and kinematics of the broad-line regions in active galaxies from IUE variability data

\bibitem[Koratkar
\& Gaskell(1989)]{1989ApJ...345..637K} Koratkar, A.~P., \& Gaskell, C.~M.\ 1989, \apj, 345, 637

\bibitem[Koratkar
\& Gaskell(1991)]{1991ApJS...75..719K} Koratkar, A.~P., \& Gaskell, C.~M.\ 1991, \apjs, 75, 719

\bibitem[Kova{\v c}evi{\'c} et al.(2010)]{2010ApJS..189...15K} Kova{\v
c}evi{\'c}, J., Popovi{\'c}, L.~{\v C}.,
\& Dimitrijevi{\'c}, M.~S.\ 2010, \apjs, 189, 15

\bibitem[Labiano(2008)]{2008A&A...488L..59L} Labiano, A.\ 2008, \aap, 488, L59


\bibitem[Lu et al.(2007)]{2007AJ....133.1615L} Lu, Y., Wang, T., Zhou, H.,
\& Wu, J.\ 2007, \aj, 133, 1615

\bibitem[Markwardt(2009)]{2009ASPC..411..251M} Markwardt, C.~B.\ 2009, in
Astronomical Data Analysis Software and Systems XVIII,
ed. D. A. Bohlender, D. Durand, \& P. Dowler (San Francisco: ASP), 251

\bibitem[Nelson \& Whittle(1996)]{1996ApJ...465...96N} Nelson, C.~H., \& Whittle, M.\ 1996, \apj, 465, 96

\bibitem[Netzer et al.(2004)]{2004ApJ...614..558N} Netzer, H., Shemmer, O.,
Maiolino, R., Oliva, E., Croom, S., Corbett, E.,
\& di Fabrizio, L.\ 2004, \apj, 614, 558

\bibitem[Richards et al.(2002)]{2002AJ....124....1R} Richards, G.~T.,
Vanden Berk, D.~E., Reichard, T.~A., Hall, P.~B., Schneider, D.~P.,
SubbaRao, M., Thakar, A.~R., \& York, D.~G.\ 2002, \aj, 124, 1

\bibitem[Richards et al.(2011)]{2011AJ....141..167R} Richards, G.~T., et
al.\ 2011, \aj, 141, 167

\bibitem[Risaliti et al.(2011)]{2011MNRAS.411.2223R} Risaliti, G., Salvati,
M., \& Marconi, A.\ 2011, \mnras, 411, 2223

\bibitem[Shen et al.(2011)]{2011ApJS..194...45S} Shen, Y., et al.\ 2011,
\apjs, 194, 45

\bibitem[Steiner(1981)]{1981ApJ...250..469S} Steiner, J.~E.\ 1981, \apj,
250, 469

\bibitem[Vasudevan \& Fabian(2009)]{2009MNRAS.392.1124V} Vasudevan, R.~V., \&
Fabian, A.~C.\ 2009, \mnras, 392, 1124

\bibitem[V{\'e}ron-Cetty et al.(2004)]{2004A&A...417..515V}
V{\'e}ron-Cetty, M.-P., Joly, M., \& V{\'e}ron, P.\ 2004, \aap, 417, 515

\bibitem[Vestergaard \& Peterson(2006)]{2006ApJ...641..689V} Vestergaard, M., \&
Peterson, B.~M.\ 2006, \apj, 641, 689

\bibitem[Wang et al.(2009)]{2009ApJ...707.1334W}
Wang, J.-G., et al. 2009, \apj, 707, 1334  %(J.G. mass formalism)

\bibitem[Wang et al.(2011)]{2011arXiv1106.2584W} Wang, H., Wang, T., Zhou,
H., Liu, B., Wang, J., Yuan, W., \& Dong, X.\ 2011, arXiv:1106.2584, accepted for publication in ApJ

\bibitem[Wang\& Lu(2001)]{2001A&A...377...52W} Wang, T., \& Lu, Y.\ 2001, \aap,
377, 52

\bibitem[Xu \& Komossa(2010)]{2010ScChG..53S.216X}
Xu, D., \& Komossa, S.\ 2010, Science in China G: Physics and Astronomy, 53, 216


\bibitem[York et al.(2000)]{2000AJ....120.1579Y} York, D.~G., et al.\ 2000,
\aj, 120, 1579

\bibitem[Zamanov et al.(2002)]{2002ApJ...576L...9Z} Zamanov, R., Marziani,
P., Sulentic, J.~W., Calvani, M., Dultzin-Hacyan, D., \& Bachev, R.\
2002, \apjl, 576, L9

\bibitem[Zhou et al.(2006)]{2006ApJS..166..128Z} Zhou, H., Wang, T., Yuan,
W., Lu, H., Dong, X., Wang, J., \& Lu, Y.\ 2006, \apjs, 166, 128



\end{thebibliography}
\end{document}